\documentclass[pra,aps,twocolumn,showpacs,superscriptaddress]{revtex4-1}
\usepackage{amssymb}
\usepackage{amsmath}
\usepackage{amstext}
\usepackage{graphicx}
\usepackage{bm}
\usepackage{booktabs}
\usepackage{dcolumn}
\usepackage{upgreek}
\usepackage{color}
\begin{document}
\title{A two-state Kalman estimator for atomic gravimetry}

\author{Bo-Nan Jiang}
\email{bnjiang@sut.edu.cn}
\affiliation{School of Science, Shenyang University of Technology, Shenyang, Liaoning 110870, China}

\date{\today}

\begin{abstract}
We present a two-state Kalman estimator of gravity acceleration and evaluate its performance by numerical simulations and post-measurement demonstration with real-world atomic gravimetry. We show that the estimator-enhanced gravimetry significantly improves upon both short-term sensitivity and long-term stability. The estimates of gravity acceleration demonstrate a $\tau^{1/2}$ feature well under white phase noise in the short term, and continue to improve as $\tau^{-1/2}$ or improve faster as $\tau^{-1}$ in the long term. This work validates the estimation of gravity acceleration as a key topic for future atomic gravimetry.
\end{abstract}
\maketitle
\section{INTRODUCTION}
Atomic gravimetry\cite{Kasevich} has proven to be a powerful tool that offers precise and accurate measurement for gravity acceleration\cite{Geiger}. It greatly complements state-of-the-art classical devices and finds applications in many fields\cite{Kai}, such as fundamental physics, metrology, inertial sensing, and geophysics. Noise from various sources limits the precision of atomic gravimetry\cite{Legouet}, including vibration noise, the phase noise of Raman lasers, detection noise, and ultimately the quantum projection noise. Relying on a combination of vibration compensation\cite{Hensley,Merlot}, low phase noise laser system\cite{Santarelli,Cacciapuoti,Jiang}, and efficient detection scheme\cite{Rocco}, residual vibration noise can be reduced down to 10 - 100 mrad per shot and other residual noise to mrad per shot level\cite{Geiger}. The residual noise level to date enables the best-reported performance of atomic gravimetry with a short-term sensitivity of 4.2 $\mu$Gal$/\sqrt{\rm{Hz}}$\cite{Hu} and a long-term stability of 0.05 $\mu$Gal\cite{Freier}. Although state-of-the-art atomic gravimeters\cite{Hu,Freier,Fu,Gillot,Menoret,Wu,Karcher} are far from approaching their intrinsic noise limit, which is set by quantum projection noise, further improvement in precision has been very challenging to achieve and has not been reported yet.

As demonstrated in previous works of atomic gravimetry, local gravity acceleration has to be estimated by averaging noisy gravimeter readings to complete gravity measurements\cite{Hu,Freier,Fu,Gillot,Menoret,Wu,Karcher}. It reveals a fact that should not be dismissed, that is, essentially, atomic gravimetry with residual noise is also a statistical problem of estimating gravity acceleration from gravimeter readings, i.e., waveform estimation\cite{Durbin}. Theoretically, one approach to waveform estimation is to build an estimator based on the state-space model\cite{Brown}. Therefore, when we cannot further reduce residual noise technically, an accurate estimation algorithm based on the state-space model that precisely describes the random process of atom interferometry becomes the key to improving the precision of atomic gravimetry.

However, the estimation algorithm for atomic gravimetry was less studied. The average method demonstrated in previous works\cite{Hu,Freier,Fu,Gillot,Menoret,Wu,Karcher} is a non-real-time algorithm without the statistical model of atomic gravimetry. As a result, rather than improving the noise characteristics of sensors, the algorithm merely provides a trade-off between white phase noise and random walk phase noise or other instrumental or geophysical effects at the cost of a non-physically contradiction between precision and bandwidth. Moreover, the algorithm treats the residual noise and gravity information equivalently, which leads to distortion or filtering out of the gravity variation with a period shorter than the measuring time. These drawbacks of the average method strongly indicate that a real-time algorithm based on a precise space-state model of atom interferometry would further benefit the precision of atomic gravimetry.

Motivated by these challenges and opportunities, we present a Kalman estimator of gravity acceleration based on a two-state model rooted in the physics of atom interferometry, which provides real-time estimation and a statistical description of noisy atomic gravimetry. The Kalman estimator is optimal for gravimetry governed by Gaussian process\cite{Kalman1,Kalman2} and is also applied for simplicity, versatility, and controllability when not optimal\cite{Kalman3,Kalman4}. We evaluate the performance of the estimator by numerical simulations and post-measurement demonstration with real-world atomic gravimetry. We show that the estimator significantly improves the short-term sensitivity and long-term stability of gravimetry by removing white phase noise in the short term and compensating random walk phase noise and other instrumental or geophysical effects in the long term. The precision limit of the estimator and the systematic error it introduces are also discussed.
\section{THE TWO-STATE KALMAN ESTIMATOR}
We briefly describe atomic gravimetry\cite{Hu,Freier,Fu,Gillot,Menoret,Wu,Karcher}. The test mass of the gravimeter is a free-falling cloud of Alkali metal atoms. To initialize the gravimetry, we trap and cool an atomic cloud to several $\mu$K in a three-dimensional magneto-optical trap and select $N$ atoms from the cloud to participate in the interferometry. The interferometry interrogates the atoms with three Raman pulses separated by two equal time intervals $T$ during the free fall, and measures the interference fringe by state-selective detection. The state-selective detection and initialization process compose dead time that interrupts the interrogation between adjacent measurements. For every sampling time of $T_s$, the interferometer cycles several shots to acquire transition probabilities of the interference fringe and reject direction-independent systematic errors by reversing the direction of the momentum transfer of Raman transitions $k_{\rm{eff}}$, and then generates one gravimeter reading $g(t)$ by the full-fringe or mid-fringe protocol.

In most previous works of atomic gravimetry\cite{Hu,Freier,Fu,Gillot,Menoret,Wu,Karcher}, the gravimeter reading is perturbed by white phase noise (WPN)
\begin{equation}
w_1(t)\sim\mathcal{N}(0,Q_1),
\end{equation}
and random walk phase noise (RWPN)
\begin{eqnarray}
\xi_2(t) &=& \int^{t}_{0}w_2(t^{'})dt^{'}, \nonumber \\
w_2(t) &\sim& \mathcal{N}(0,Q_2),
\end{eqnarray}
or other instrumental or geophysical effects, such as fluctuation of systematic errors, imperfect tide model, ocean loading, etc. Here, the random process $w_{1,2}(t)$ is Gaussian with zero mean and variance $Q_{1,2}$, and $\xi_2(t)$ denotes phase errors accumulating with a random walk.

Considering the experimental fact above, we choose to model the state of the atomic gravimetry with a two-state vector
\begin{equation}
  x(t) =
  \begin{bmatrix}
  x_{1}(t) \\
  x_{2}(t)
  \end{bmatrix},
\end{equation}
where $x_{1}(t)$ is an integration of gravimeter readings
\begin{eqnarray}
  x_{1}(t) &=& \int^{t}_{0}g(t^{'})dt^{'} \nonumber \\
  &=& \int^{t}_{0}g_0(t^{'})dt^{'}+\int^{t}_{0}w_1(t^{'})dt^{'}+\int^{t}_{0}\xi_2(t^{'})dt^{'},
\label{eq:x1}
\end{eqnarray}
$x_{2}(t)$ describes the cumulative phase error
\begin{equation}
  x_{2}(t)=\xi_2(t),
\end{equation}
and $g_0(t)$ is the true absolute value of gravity acceleration.

As expected, quantum projection noise\cite{Itano} sets the intrinsic limit of the model to resolve the state. It disturbs each measurement by $w_1(t)$, the variance of which is evaluated as
\begin{equation}
  Q_1=\left(\frac{1}{k_{\rm{eff}}T^2}\frac{1}{\sqrt{N}}\right)^2,
\end{equation}
and also generates a random phase error $w_2(t)$ between adjacent measurements, the variance of which is evaluated as
\begin{equation}
  Q_2=\left(\frac{1}{k_{\rm{eff}}T^2}\frac{1}{\sqrt{N}}\right)^2\frac{1}{T^2_s}.
\end{equation}

The state is measured at a sequence of nominal times $n$, during which we reasonably treat $g_0(n)$ as cyclostationary, as the periods of the gravity variation are much longer than $T_s$. The model for the discrete-time process of $x(n)$ is set up as follows:
\begin{equation}
  x(n) = Fx({n-1}) +
  \begin{bmatrix}
  g_0(n)T_s \\
  0
  \end{bmatrix}
  + W(n),
\label{eq:two-state model}
\end{equation}
where the state transition matrix
\begin{equation}
  F = \begin{bmatrix}
  1 & T_s \\
  0 & 1
  \end{bmatrix} \nonumber,
\end{equation}
$W(n)$ represents the random process
\begin{equation}
  W(n)=\int^{t(n)}_{t(n-1)}\begin{bmatrix}
  1 & t(n)-t^{'} \\
  0 & 1
  \end{bmatrix} \nonumber \begin{bmatrix}
  w_1(t^{'}) \\
  w_2(t^{'})
  \end{bmatrix} \nonumber dt^{'},
\end{equation}
and the covariance matrix $Q$ for $W(n)$ is given by
\begin{equation}
  Q = E[W(n) W(n)^T] = \begin{bmatrix}
  Q_1T_s+Q_2\frac{T_s^3}{3} & Q_2\frac{T_s^2}{2} \\
  Q_2\frac{T_s^2}{2} & Q_2T_s
  \end{bmatrix} \nonumber.
\end{equation}

In the framework of Kalman estimator, we estimate $\widehat{x}(n)$ and finally $\widehat{g}_0(n)$ by a Kalman recursion based on the two-state model above in two steps:

In the propagation step, the a priori estimates are predicted conditioned on the previous a posteriori estimates as follows
\begin{eqnarray}
\widehat{x}^-(n) &=& F \widehat{x}({n-1}) + \begin{bmatrix}
  \widehat{g}_0^-(n)T_s \\
  0
  \end{bmatrix}, \nonumber \\
P^-(n) &=& F P({n-1})F^T + Q,
\end{eqnarray}
where the superscript $"-"$ indicates the a priori estimate deduced according to the model, $P$ is the error covariance matrix, and $\widehat{g}_0^-(n)$ is generated with the theoretical prediction of solid-earth tide\cite{Dehant} and an a priori short-term measurement of the local gravity.

Then, in the update step, the a priori estimates are refined after each observation constructed recursively $z(n) = z({n-1}) + g(n)T_s$, and the a posteriori estimates are deduced according to
\begin{eqnarray}
\widehat{x}(n) &=& \widehat{x}^-(n) + K(n) [z(n)-H\widehat{x}^-(n)], \nonumber \\
P(n) &=& [I-K(n)H]P^-(n), \nonumber
\end{eqnarray}
where the measurement matrix
\begin{equation}
  H = \begin{bmatrix}
  1 & 0 \\
  \end{bmatrix} \nonumber,
\end{equation}
the Kalman gain
\begin{equation}
  K(n) = P^-(n)H^T [HP^-(n)H^T + (n+1)RT_s^2]^{-1},
\end{equation}
and $R$ is experimentally determined from the variance of a sample of $g(n)$.

The Kalman estimator is initialized according to our a prior knowledge of the atomic gravimeter, with
\begin{eqnarray}
\widehat{x}^-(0) &=& \begin{bmatrix}
  g(0)T_s \\
  \sqrt{Q_2}T_s
  \end{bmatrix}, \nonumber \\
P^-(0) &=& Q, \nonumber \\
z(0) &=& g(0)T_s.
\end{eqnarray}

Finally, the estimate $\widehat{g}_0(n)$ is derived from the estimate $\widehat{x}$ through the following equation
\begin{equation}
  \widehat{g}_0(n) = \frac{\widehat{x}_{1}(n)-\widehat{x}_{1}({n-1})}{T_s}-\widehat{x}_{2}(n),
\label{eq:est_g0}
\end{equation}
where the first term $\frac{\widehat{x}_{1}(n)-\widehat{x}_{1}({n-1})}{T_s}$ unveils the gravity acceleration beneath the white phase noise in the short term, and the second term $\widehat{x}_2$ monitors and compensates the cumulative phase error in the long term. Therefore, the estimate $\widehat{g}_0(n)$ assuages both white phase noise and random walk phase noise in atomic gravimetry.

We evaluate the precision limit of the Kalman estimator by investigating the theoretical prediction of the Allan deviation of the estimate $\widehat{g}_0$, that is $\sigma_{E}$. In the short term, $\sigma_{E}$ is determined by the intrinsic resolution limit of the model
\begin{equation}
  \sigma_E=\sqrt{\frac{Q_1}{\tau}+\frac{Q_2\tau}{3}},
\end{equation}
which approximately scales as $\tau^{1/2}$ after $\tau_1\simeq\sqrt{\frac{3Q_1}{Q_2}}$. Obviously, the crossover between $\sigma_{E}$ and the white phase noise observed $\sqrt{\frac{Q_W}{\tau}}$ inevitably occurs at
\begin{equation}
  \tau_2\simeq\sqrt{\frac{3Q_W}{Q_2}}.
\end{equation}
Beyond $\tau_2$, as $\sqrt{\frac{Q_W}{\tau}}$ dives under $\sqrt{\frac{Q_1}{\tau}+\frac{Q_2\tau}{3}}$, the noise of the estimate $\widehat{g}_0$ is then predicted to be dominated by the white phase noise observed
\begin{equation}
  \sigma_E=\sqrt{\frac{Q_W}{\tau}},
\end{equation}
and improves further as $\tau^{-1/2}$. Moreover, when the long-term measurement is dominated by the white phase noise, as the ultrastable atomic gravimetry reported by Ref. \cite{Freier}, $\widehat{x}_{2}$ in Eq. \ref{eq:est_g0} effectively correlates the adjacent measurements such that the phase error is stationary\cite{Biedermann}. Then we find
\begin{equation}
  \sigma_E=\sqrt{\frac{\frac{\tau}{T_s}Q_W}{\frac{\tau^2}{T_s}}} \longrightarrow \sqrt{\frac{Q_W}{\frac{\tau^2}{T_s}}},
\end{equation}
and improves as $\tau^{-1}$ after $\tau_2$. We also consider $\sigma_E$ as a systematic error introduced by the Kalman estimator. The ratio of $\sigma_E$ to Type A uncertainty of the gravimetry under identical conditions is estimated to be
\begin{equation}
r\simeq
\begin{cases}
\sqrt{\frac{Q_2}{3Q_W}}\tau, & \text{WPN observed at } \tau<\tau_2,\\
\sqrt{\frac{3Q_W}{Q_{RW}}}\frac{1}{\tau}, & \text{RWPN observed at }  \tau>\tau_2,\\
\sqrt{\frac{T_s}{\tau}}, & \text{WPN observed at }  \tau>\tau_2.
\end{cases}
\end{equation}
Then, the systematic error introduced by $\sigma_E$ is reasonably neglectable in the short term before $\frac{\tau_2}{10}$ or in the long term beyond $10\tau_2$, as the ratio in this regime
\begin{equation}
r\simeq
\begin{cases}
<\frac{1}{10}, & \text{WPN observed at } \tau<\frac{\tau_2}{10},\\
<\sqrt{\frac{Q_2}{Q_{RW}}}\frac{1}{10}\sim\frac{1}{10}, & \text{RWPN observed at }  \tau>10\tau_2,\\
<\sqrt{\frac{T_s}{10\tau_2}}\ll\frac{1}{10}, & \text{WPN observed at }  \tau>10\tau_2.
\end{cases}
\end{equation}
Meanwhile, between $\frac{\tau_2}{10}$ and $10\tau_2$, $r\sim 1$, which is also an acceptable level of accuracy\cite{Karcher}.
\begin{figure}[tbp]
\includegraphics[width=0.5\textwidth]{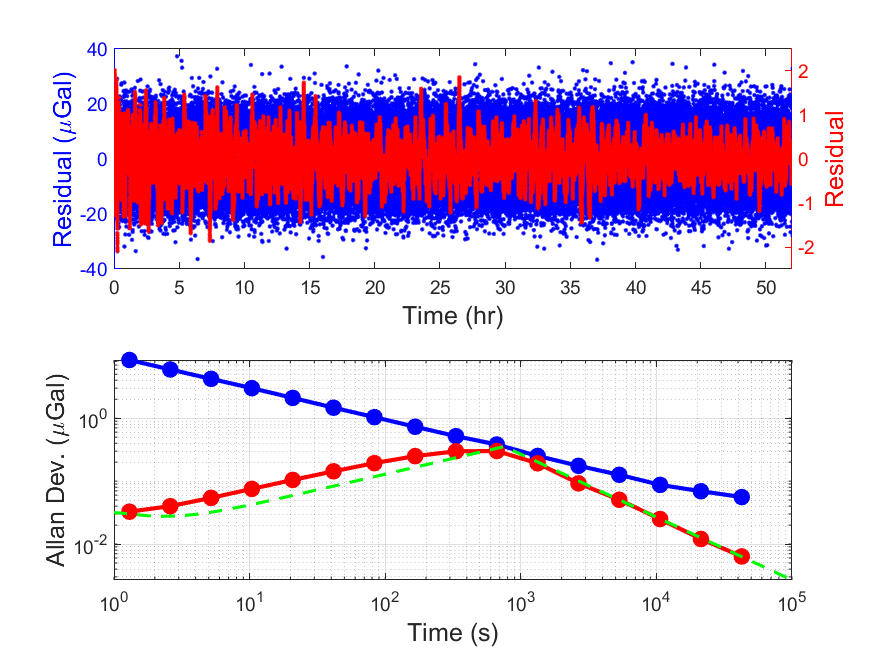}
\caption{(Color online) Top: Residual acceleration by subtracting $g^{\mathrm{sim}}_0(n)$ from the the simulated gravimeter readings of Set One (blue) or the Kalman estimates (red). Bottom: Allan deviation calculated from the residual acceleration. The green dashed line is the precision limit of the Kalman estimator.}
\label{fig:simw}
\end{figure}

\section{NUMERICAL RESULTS}
In numerical simulations, we estimate gravity acceleration from the simulated gravimeter readings. We generate these gravimeter readings $g^{\mathrm{sim}}(n)$ by the two-state model in Eq. \ref{eq:two-state model}, where the true value of the gravity acceleration is simulated by the theoretical prediction of solid-earth tide and denoted as $g^{\mathrm{sim}}_0(n)$. And we evaluate the performance of the Kalman estimator by investigating the residual acceleration and the corresponding Allan deviation.

We generate two sets of simulated gravimeter readings according to the best-reported performance of atomic gravimetry\cite{Hu,Freier} to push the estimator to the limit and evaluate its performance. The data of Set One\cite{Freier} show ultrahigh stability with white phase noise $Q_W=(9.6$ $\mu$Gal$/\sqrt{\mathrm{Hz}})^2$ that integrates as $\tau^{-1/2}$ for a time scale of $10^5$ s. $N\sim10^7$ $^{87}\mathrm{Rb}$ atoms participate in the interferometry, being interrogated by a 780nm pulse sequence with $T=260$ ms. The sampling time $T_s$ for Set One $\sim 1.3$ s. Meanwhile, the data of Set Two\cite{Hu} show ultrahigh sensitivity with white phase noise $Q_W=(4.2$ $\mu$Gal$/\sqrt{\mathrm{Hz}})^2$ and random walk phase noise $Q_{RW}\sim3\times(0.01$ $\mu$Gal$/\sqrt{\mathrm{s}})^2$. $N\sim5\times10^7$ $^{87}\mathrm{Rb}$ atoms participate in the interferometry, being interrogated by a 780nm pulse sequence with $T=300$ ms. The sampling time $T_s$ for Set Two is $2$ s.
\begin{figure}[tbp]
\includegraphics[width=0.5\textwidth]{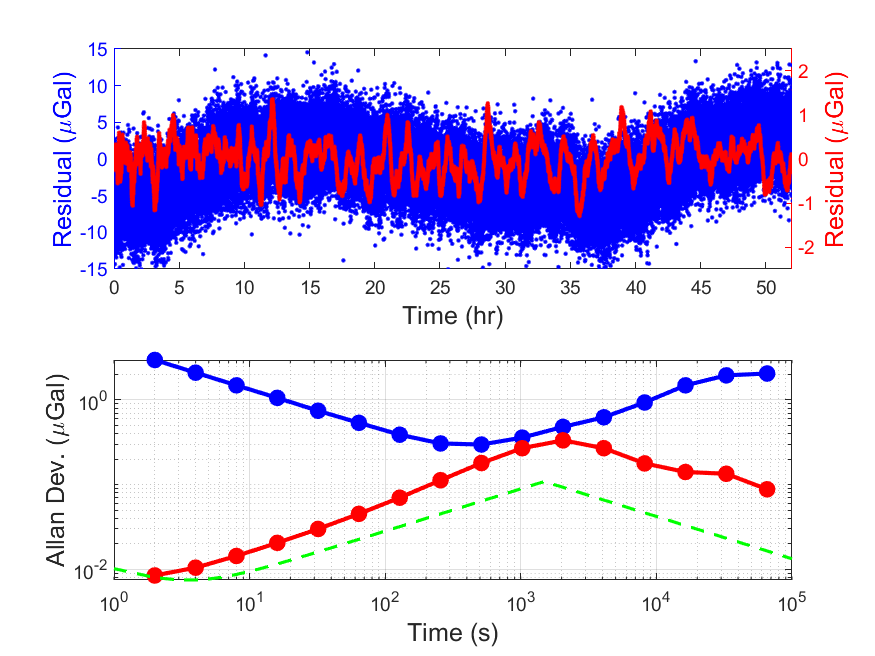}
\caption{(Color online) Top: Residual acceleration by subtracting $g^{\mathrm{sim}}_0(n)$ from the the simulated gravimeter readings of Set Two (blue) or the Kalman estimates (red). Bottom: Allan deviation calculated from the residual acceleration. The green dashed line is the precision limit of the Kalman estimator.}
\label{fig:simwrw}
\end{figure}

Fig. \ref{fig:simw} shows the numerical results of estimating gravity acceleration from the data of Set One, where the white phase noise dominates the gravimetry. Compared to the simulated gravimeter readings, the estimates agree better with $g^{\mathrm{sim}}_0(n)$, as the residual acceleration demonstrates a zero mean and a standard deviation improved by a factor of $\frac{8.42\thinspace\mu\mathrm{Gal}}{0.43\thinspace\mu\mathrm{Gal}}\simeq 20$. The Allan deviation of the estimates integrate as nearly $\tau^{1/2}$ in the short term and improves as $\tau^{-1}$ in the long term, which is consistent with the theoretical prediction. The numerical results given by Allan deviation also hightlight a short-term sensitivity of 0.02 $\mu$Gal$/\sqrt{\rm{s}}$ and a long-term stability of 0.006 $\mu$ Gal at a measurement time of $0.4\times10^5$ s.

Fig. \ref{fig:simwrw} shows the numerical results of estimating gravity acceleration from the data of Set Two, where the white phase noise and the random walk phase noise govern the gravimetry. It is found that though the simulated gravimeter readings are zigzagged from $\sim$-15 $\mu$Gal to $\sim$10 $\mu$Gal, the estimates agree well with $g^{\mathrm{sim}}_0(n)$, demonstrating a zero mean and a standard deviation of 0.38 $\upmu\rm{Gal}$. The Allan deviation of the estimates improves as $\tau^{-1/2}$ in the long term and reaches a stability of 0.087 $\upmu\rm{Gal}$ at a measurement time of $0.65\times10^5$ s, whereas the stability of simulated data degrades as $\tau^{1/2}$ after reaching 0.29 $\mu\rm{Gal}$ in the mid term. The Kalman estimator also improves the short term sensitivity from $4.2$ $\mu$Gal$/\sqrt{\mathrm{Hz}}$ to 0.007 $\mu$Gal$/\sqrt{\rm{s}}$.

The numerical results demonstrates that the two-state Kalman estimator not only remove the white phase noise in the short term, but also compensates the cumulative phase error in the long term and makes the Allan deviation of the estimator-enhanced gravimetry continue to improve or improve faster with measurement time, which could provide a significant advantage in both short-term and long-term measurements. Moreover, the good agreement between the Kalman estimates and $g^{\mathrm{sim}}_0(n)$ shows a reliable method that will not obscure the phenomena observed with nontrivial mathematically-originated bias.

We also briefly discuss the estimator's response to acceleration jumps. The limit to the detectable amplitude of the acceleration jump is evaluated by six times of the standard deviation of the residual acceleration, which is the difference needed to completely separate the two quantities that are perturbed by random noises. Based on the numerical results of Set One and Set Two, this limit is calculated to be 2.58 $\mu$Gal and 2.28 $\mu$Gal respectively. The bandwidth of the Kalman estimator is evaluated by the period of generating one $\widehat{g}_0$, which consists of the computing time $T_{\rm{cal}}$ and the time $T_s$ that generates one observation to update the a priori estimate,
\begin{equation}
  BW=\frac{1}{T_{\rm{cal}}+T_s}.
\end{equation}
$T_{\rm{cal}}$ is measured to be 2 - 4 $\mu$s by a timer set in codes, and as $T_s\gg T_{\rm{cal}}$, the bandwidth of the estimator is significantly limited by the sampling rate of the sensor
\begin{equation}
  BW\simeq\frac{1}{T_s}.
\end{equation}
For the best-reported performance of atomic gravimetry referenced by the numerical simulations, the bandwidth of the estimator is calculated to be 0.77 Hz\cite{Freier} and 0.5 Hz\cite{Hu}, whereas for the high-data-rate sensor aiming at inertial navigation, the bandwidth is calculated to be 48 Hz to 248 Hz\cite{McGuinness}.
\section{POST-MEASUREMENT DEMONSTRATION}
\begin{figure}[tbp]
\includegraphics[width=0.5\textwidth]{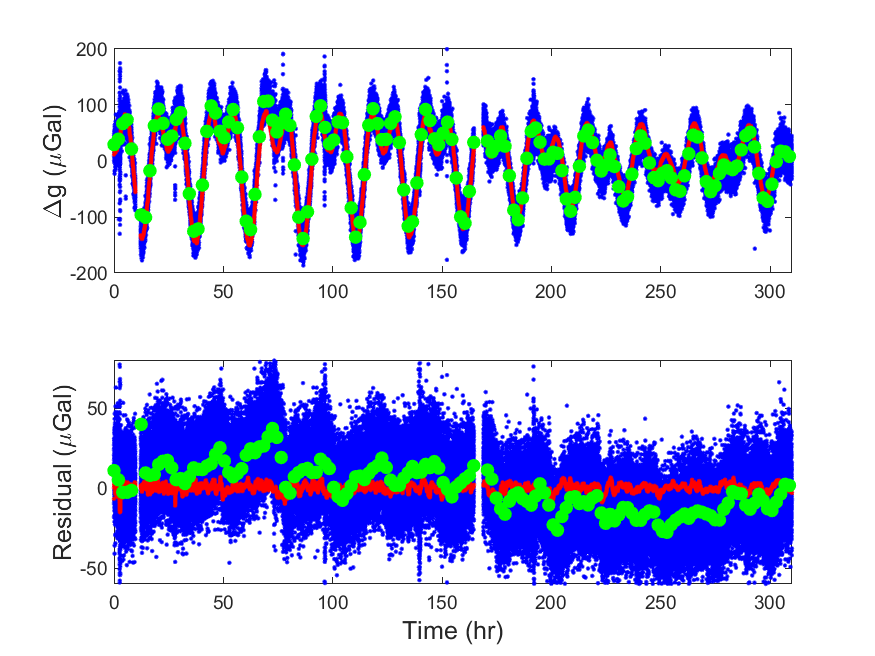}
\caption{(Color online) Long-term measurement of gravity for 310 hours and the corresponding residual acceleration by subtracting
the solid-earth tide from the gravimeter readings (blue) or the Kalman estimates (red). The green dot is obtained by averaging the gravimeter readings over every 2 hours.}
\label{fig:exp_res}
\end{figure}
\begin{figure}[tbp]
\includegraphics[width=0.5\textwidth]{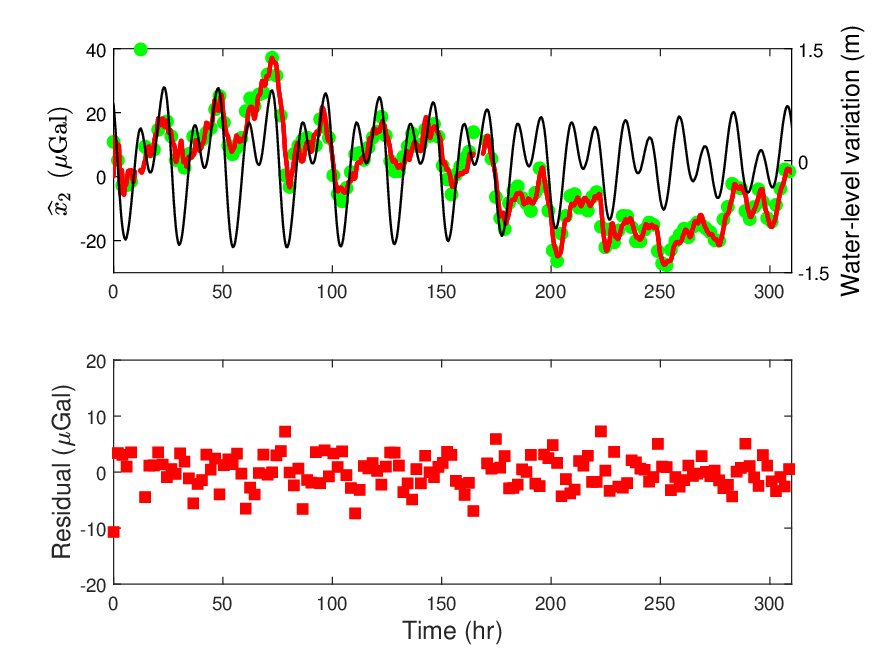}
\caption{(Color online) The estimates $\widehat{x}_2$ (red) and the residual acceleration by subtracting the residual of the 2-hour average (green) from $\widehat{x}_2$. The water-level variation (black) is measured by the observatory of National Oceanic and Atmospheric Administration in Richmond, California.}
\label{fig:exp_x2}
\end{figure}
\begin{figure}[tbp]
\includegraphics[width=0.5\textwidth]{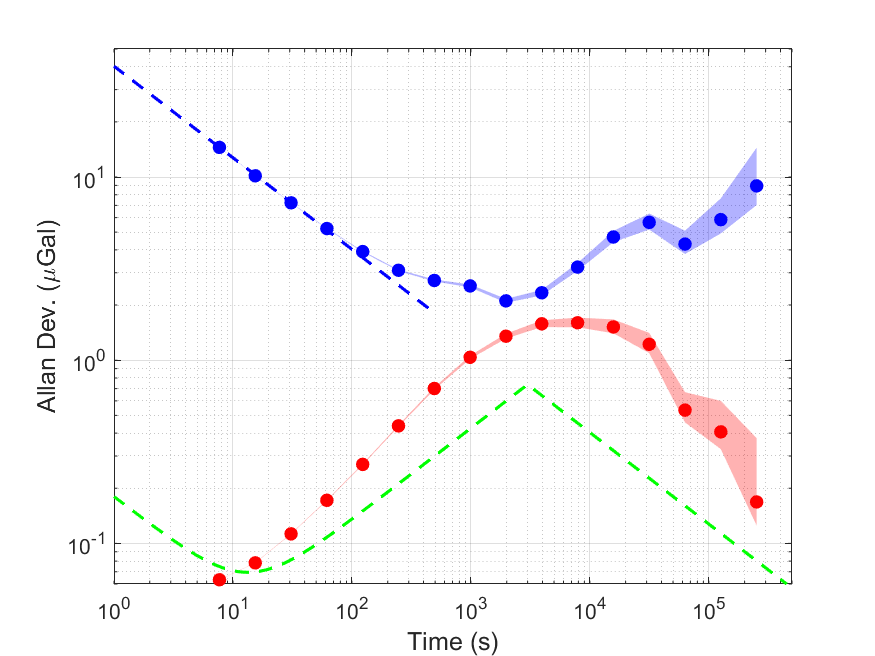}
\caption{(Color online) Allan deviation of the gravity measurement calculated from the residual acceleration of 310 hours: blue dot, gravimeter readings; red dot, Kalman estimates. The blue dashed line corresponds to a sensitivity of $40.3$ $\mu$Gal$/\sqrt{\mathrm{Hz}}$. And the green dashed line is the precision limit of the Kalman estimator.}
\label{fig:exp_allan}
\end{figure}

To evaluate the performance of the Kalman estimator thoroughly, we further performs a post-measurement demonstration by applying the estimator to the real-world atomic gravimetry in Ref. \cite{Wu}, where Wu et al. operated a mobile atomic gravimeter at their laboratory in Berkeley, and performed a long-term measurement of the local gravity from January 2, 2019 to January 15, 2019. These data were chosen because they consisted of a fairly long record with sufficient complicated noise characteristics. The outliers in raw data are mainly brought by the vibration noise, and are kept in the time series to test the robustness of the estimator. All experimental data of raw gravimeter readings were provided by the authors of Ref. \cite{Wu} and used with permission.

We contend that though being applied to the post-measurement demonstration in this work, the intrinsic characteristic of the Kalman algorithm makes it easy to be generalized to the real-time implementation.

As shown in Fig. \ref{fig:exp_res}, the mobile atomic gravimeter that is described in Ref. \cite{Wu} performs a continuous measurement of the local gravity for 310 hours. The test mass of the gravimeter is a free-falling cloud of cold cesium atoms. $N\sim5\times10^6$ atoms participate in the interferometry, being interrogated by a 852nm pulse sequence with $T=130$ ms. For every $T_s=7.696$ s, the interferometer cycles 16 shots and generates one gravimeter reading. Compared to the raw gravimeter readings (blue) and the data averaged over every 2 hours (green), the Kalman estimates (red) agrees well with the solid-earth tide, demonstrating residual acceleration with a zero mean and a standard deviation improved by a factor of $\sim 10$. Meanwhile, the residual acceleration of both the raw data and the 2-hour average clearly shows nontrivial long-term gravity variation composed of multiple components with various periods.

In Fig. \ref{fig:exp_x2}, comparison between the residual of the 2-hour average (green) and the water-level variation (black) shows that one component of the long-term gravity variation is correlated to ocean loading, which is consistent with Ref. \cite{Wu}. However, with other components in residual being unexplained, especially the obvious long-term drift, it indicates that the long-term performance of the gravimeter is limited by complicated effects, such as cumulative phase errors, fluctuation of systematic errors, imperfect tide model, ocean loading, etc. Though complicated, the estimate $\widehat{x}_2$ (red) faithfully monitors the long-term gravity variation and compensates it in the estimate $\widehat{g}_0$. As no superconducting gravimeter provides true gravity variation, the estimation error of $\widehat{x}_2$ is evaluated by subtracting the residual acceleration of the 2-hour average from $\widehat{x}_2$. The residual demonstrates a zero mean and a standard error of 3.39 $\mu$Gal that is in good agreement with  $\sqrt{\rm{ADev}(g_{2hr})^2+\rm{ADev}(\widehat{g}_{7.696s})^2}=\sqrt{(3.39 \mu \rm{Gal})^2+(0.06 \mu \rm{Gal})^2}\simeq3.39$ $\mu$Gal. Obviously, $\widehat{x}_2$ monitors the long-term gravity variation with a precision better than the average method, as the measurement noise of the gravimeter dominates the estimation error. The results in Fig. \ref{fig:exp_x2} also demonstrate that the Kalman estimator is very robust to the long-term gravity variation, it not only corrects the cumulative phase error that is conceived in the model, but also is a very powerful solution to other sources of variation, such as ocean loading. This feature is particularly attractive for gravity survey\cite{Hu,Wu} or comparison event\cite{Shuqing}, where cumulative phase errors, fluctuation of systematic errors, and imperfect tide model are notorious issues that significantly degrade the performance of gravimeters.

In Fig. \ref{fig:exp_allan}, Allan deviation (ADev) of the residual acceleration is calculated to characterize the short-term sensitivity and the long-term stability of the gravimeter and the Kalman estimates. The sensitivity of the gravimeter (blue) follows $40.3$ $\mu$Gal$/\sqrt{\mathrm{Hz}}$ for up to 100 s. The noise in this regime is dominated by white phase noise, and demonstrates a reduction characteristic of $\tau^{-1/2}$ via integrating in time. For long-term stability, the ocean loading leads to the broad peak around $3\times10^4$ s, and the degradation of the stability beyond $3\times10^4$ s might be due to random walk phase noise, or other instrumental or geophysical effects. The Kalman estimator reshapes the residual noise by removing the white phase noise in the short term and compensating the long-term gravity variation. In the short term, the residual of the estimates (red) gradually increases from the precision limit (green) 0.013 $\mu$Gal$/\sqrt{\rm{s}}$  to 0.036 $\mu$Gal$/\sqrt{\rm{s}}$ around $1\times10^3$ s, and scales as nearly $\tau^{1/2}$ well under the white phase noise; In the long term, the vanishing of the broad peak and the degradation mentioned above shows that ocean loading and cumulative phase error (or other instrumental or geophysical effects) are also significantly compensated by the Kalman estimator. The Allan deviation of the Kalman estimates highlights a short-term sensitivity better than 0.036 $\mu$Gal$/\sqrt{\rm{s}}$ and a long-term stability improved by a factor of $\frac{8.95\thinspace\mu\mathrm{Gal}}{0.17\thinspace\mu\mathrm{Gal}}\simeq 52$ at a measurement time of $2.5\times10^5$ s.
\section{CONCLUSION}
In conclusion, we shed light on the two-state Kalman estimator that is rooted in the physics of atom interferometry. We evaluate the performance of the estimator by both numerical simulations and post-measurement demonstration with real-world atomic gravimetry. We show that the estimator significantly improves the short-term sensitivity and long-term stability of gravimetry by removing white phase noise in the short term and compensating random walk phase noise and other instrumental or geophysical effects in the long term. As well as the good agreement between the estimates and the solid-earth tide, the noise of the estimator-enhanced gravimetry demonstrates a $\tau^{1/2}$ feature in the short term governed by white phase noise, and continue to improve as $\tau^{-1/2}$ or improve faster as $\tau^{-1}$ in the long term dominated by random walk phase noise and other long-term effects or white phase noise respectively. We also provides a tool to evaluate the precision limit of the Kalman estimator and the systematic error it introduces by calculating the theoretical prediction of the Allan deviation of the estimates. By comparing to Type A uncertainty of the gravimetry, we find that the systematic error introduced by the estimator is acceptable for state-of-the-art atomic gravimeters.

The Kalman estimator demonstrated could provide a significant advantage in both short-term and long-term measurements, for example, reducing the time cost in gravity survey with mobile gravimeter\cite{Wu} or compensating the long-term variation of systematic errors in comparison event\cite{Xie}. This demonstration would be of great interest for those applications involving static measurements of
gravity, such as metrology or geophysics.

\section*{Acknowledgments}
We would like to thank Prof. Holger M\"{u}ller for providing the raw experimental data used in post-measurement demonstration. This work is funded by the Scientific Research Project of the Educational Department of Liaoning Province (Grant No. LJKMZ20220483) and the Youth Program of National Natural Science Foundation of China (Grant No. 11804019).

\section*{Data Availability}
The data that support the findings of this study are available from the corresponding author upon reasonable request.

\end{document}